\documentclass{appolb}
\usepackage{epsfig}
\def\1{\mbox{l\hspace{-0.53em}1}}

\begin{document}
\title{Lowest negative parity baryons in the $1/N_c$ expansion
\thanks{Presented by N. Matagne at the EMMI Workshop and Max-Born Symposium, Wroclaw, July 9--11, 2009.}%
}
\author{N. Matagne\footnote{F.R.S.-FNRS Postdoctoral Research Fellow; E-mail: nicolas.matagne@umons.ac.be}
\address{
Groupe de Physique Nucl\'eaire et Subnucl\'eaire, University of Mons, 
Place du Parc 20, B-7000 Mons, Belgium}
\and
Fl. Stancu\footnote{E-mail: fstancu@ulg.ac.be}
\address{University of Li\`ege, Institute of Physics B5, Sart Tilman, B-4000 Li\`ege 1, Belgium}
}
\maketitle
\begin{abstract}
We review a recently proposed approach to study the lowest negative 
parity baryons within the $1/N_c$ expansion. The method is based on
the derivation of the matrix elements of SU(2N$_f$) generators 
for mixed symmetric $[N_c-1,1]$ flavor-spin states. Presently it is applied 
to the N = 1 band and a comparison is made with a former method based 
on the decoupling of the system into a  symmetric core of $N_c-1$ quarks 
and an excited one. We prove that the decoupling is not necessary
and moreover, it misses some important physical consequences. 
\end{abstract}
\PACS{12.39.-x; 11.15.Pg; 11.30.Hv}
  
\section{Introduction}
In the energy regime of  the hadron spectroscopy, QCD does not admit a 
perturbative expansion in terms of the coupling constant of the theory. 
A interesting solution to the problem is to generalize QCD to $N_c$ colors, 
$1/N_c$ becoming the expansion parameter of the theory \cite{tHo74,Wit79}. 
In the large $N_c$ limit, QCD possesses an exact contracted SU$(2N_f)_c$ 
symmetry \cite{Gervais:1983wq,DM93}. For large but finite $N_c$, the contracted 
SU$(2N_f)_c$  becomes the SU($2N_f$) symmetry group of the constituent quark model. 
The ground states baryons are degenerate in the $N_c \to \infty$ limit and 
the mass splitting starts at order $1/N_c$ \cite{DJM94,DJM95}.

Following the quark model classification
of excited baryons, it is generally assumed that in the large 
$N_c$ limit, the SU($2N_f)\times$O(3) symmetry is satisfied. 
For the $N = 1$ band, at large but finite $N_c$,
it has been shown that this symmetry is broken at order $N_c^0$
by several operators, in particular by the spin-orbit operator,
so that the mass splitting starts at order  $N_c^0$.  

This review is devoted to the description of the lowest  negative 
parity baryons, belonging to the $[{\bf 70},1^-]$ multiplet (the $N = 1$ band).
The wave function of these baryons is mixed symmetric in the orbital and
the spin-flavor parts. As a consequence, the calculation of
the spectrum becomes more complicated than for symmetric states.
During the last four years, 
this became a controversial subject. The traditional procedure was based 
on a Hartree approximation, the baryon being described by an orbitally excited 
($\ell = 1$) quark moving in the collective potential generated by a ground 
state core composed of $N_c -1$ quarks. Accordingly, only the  $s^{N_c-1}p$ 
configuration was considered in the orbital part. The other 
components of the wave function needed to obtain a mixed symmetric orbital 
and spin-flavor states were neglected  \cite{CCGL}.  Accordingly, the 
generators of SU($2N_f$) were written as the sum of two terms, one 
acting on the core and the other on the excited quark. In this way
the number of linearly independent terms in the mass formula becomes
tremendously large which creates difficulties in choosing the dominant ones.

One can go beyond this approximation and treat the wave function
of the baryons exactly in order to avoid the decoupling of the
system into a core and an excited 
quark \cite{MS1}. In this case the wave function has the correct symmetry 
but it requires the knowledge of the matrix elements of the spin-flavor 
generators  for the mixed symmetric states.  These matrix elements have been 
derived for SU(4) in the sixties \cite{HP} but only recently 
for SU(6) \cite{MS2}.

For a comparison, one can include the neglected 
terms of the wave function in the traditional decoupling picture, but 
others group theoretical difficulties appear related to the knowledge of 
isoscalar factors of the permutation group S$_{N_c}$. This problem  
has already been solved for nonstrange baryons \cite{MS3}. 
The strange baryons are underway and will be presented in a future 
publication \cite{MS4}.

In the following we shall briefly describe the two procedures.

\section{The wave function}

To satisfy the Pauli principle, the wave function describing a baryon must 
be antisymmetric. This wave function is usually the product of its orbital (O), 
spin (S), flavor (F) and color (C) parts. As in nature the baryons are 
colorless, the color part is always antisymmetric, the orbital-spin-flavor 
is then symmetric. As the color plays no role in the mass analysis, 
one integrates it out and considers only the orbital-spin-flavor part only.

Here we are interested in the multiplet $[{\bf 70},1^-]$.
Its orbital 
and spin-flavor parts must have both a mixed symmetry 
which for arbitrary $N_c$ has the partition $[N_c-1,1]$. 
In terms of inner products of the permutation group $S_{N_c}$, the wave 
function takes the form
\begin{equation}
\label{EWF}
|[N_c]1 \rangle = \frac{1}{\sqrt{N_c-1}}
\sum_{Y} |[N_c-1,1] Y \rangle_{O}  |[N_c-1,1] Y \rangle_{FS},
\end{equation}
where $Y$ is the corresponding Young tableau. Here we sum over the $N_c-1$ 
possible standard Young tableaux. The factor $1/\sqrt{N_c-1}$ represents 
the Clebsch-Gordan coefficient (CG) of $S_{N_c}$ needed to construct a symmetric 
wave function $[N_c]$ from its mixed symmetric parts. 

Let us first introduce the global wave function, without the decoupling of the excited quark from the ground state core. In that case, as the matrix elements of the operators are identical for all $Y$ due to Weyl's duality between a linear group and a symmetric group
in a given tensor space\footnote{see Ref. \cite{book}, Sec 4.5.}, one does not need to specify $Y$.  
Then the explicit form of a  wave function of total angular momentum 
$\vec{J} = \vec{\ell} + \vec{S}$ and  isospin $I$ is

\begin{eqnarray}\label{WF}
\lefteqn{|\ell SJJ_3; I I_3 \rangle  = } \nonumber \\
& & \sum_{m_\ell,m_s}
      \left(\begin{array}{cc|c}
	\ell    &    S   & J   \\
	m_\ell  &    m_s  & J_3 
      \end{array}\right) 
|[N_c-1,1]\ell m_{\ell} \rangle
|[N_c-1,1] S m_s I I_3 \rangle,
\end{eqnarray}
each term containing an SU(2) CG coefficient, an 
orbital part 
$|[N_c-1,1]\ell m_{\ell} \rangle$ an a spin-flavor
part $|[N_c-1,1] S m_s I I_3 \rangle$.

The exact but decoupled 
wave function reads \cite{MS3}
\begin{eqnarray}
 \lefteqn{|\ell S J J_3; II_3\rangle =} \nonumber \\
& & \sum_{p, p', p'', \ell_c, \ell_q, m_\ell, m_q,\atop   m_c,m_s, m_1, m_2, i_1, i_2} a(p,\ell_c,\ell_q) \left(\begin{array}{cc|c}
	\ell_c    &  \ell_q   & \ell   \\
	m_c  &    m_q    & m_\ell 
      \end{array}\right) 
   \left(\begin{array}{cc|c}
	\ell    &    S   & J   \\
	m_\ell  &    m_s  & J_3 
      \end{array}\right) \nonumber \\
& & \times K([f']p'[f'']p''|[N_c-1,1]p) 
\left(\begin{array}{cc|c}
	S_c    &    \frac{1}{2}   & S   \\
	m_1  &         m_2        & m_s
      \end{array}\right)
 \left(\begin{array}{cc|c}
	I_c    &    \frac{1}{2}   & I   \\
	i_1    &       i_2        & I_3
      \end{array}\right) \nonumber \\
& & \times |\ell_c m_c\rangle  |S_cm_1\rangle|I_ci_1\rangle |\ell_q m_q\rangle |1/2m_2\rangle 
|1/2i_2\rangle,
\label{decex}
\end{eqnarray}
where $\ell_c$ and $\ell_q$ represent the angular momenta of the core and of 
the decoupled quark respectively and where $a(p,\ell_c,\ell_q)$ are the 
one-body fractional parentage coefficients to decouple the $N_c$th quark
from the rest in the orbital part. These are given by \cite{MS3}
\begin{eqnarray}
a(2,\ell_c =0, \ell_q = 1) & = & \sqrt{\frac{N_c-1}{N_c}}, \label{2coreground}\\
a(2,\ell_c =1, \ell_q = 0) & = & -\sqrt{\frac{1}{N_c}}, \label{2coreex} \\
 a(1,\ell_c =1, \ell_q = 0) & = & 1.
\end{eqnarray}
The isoscalar factors $K([f']p'[f'']p''|[N_c-1,1]p)$ used in 
Eq. (\ref{decex}) are also given in \cite{MS3}.

\section{The mass operator}

The mass operator $M$ is defined as a linear combination of independent operators $O_i$
\begin{equation}\label{BMASS}
 M = \sum_{i} c_i O_i,
\end{equation}
where the coefficients $c_i$ are reduced matrix elements that encode the QCD dynamics and are determined from a fit to the existing experimental data. The building blocks of the operators $O_i$ are the SU($2N_f$) generators $S_i$, $T_a$ and $G_{ia}$ and the SO(3) generators $\ell_i$. Their general form is
\begin{equation}
 O_i = \frac{1}{N_c^{n-1}}O_\ell^{(k)}\cdot O_{SF}^{(k)},
\end{equation}
where $O_\ell^{(k)}$ is a $k$-rank tensor in SO(3) and 
$O_{SF}^{(k)}$ a $k$-rank tensor in SU(2)-spin, but invariant in SU($N_f$). 
$O_i$, with $i$ a numbering index, is then rotational invariant. For the ground state one has $k=0$. 
The excited states also require $k=1$ and $k=2$ terms.


\section{Results}

\subsection{The exact wave function}

\begin{table}[h!]
\caption{List of operators contributing to the mass (\ref{BMASS}) up to order $1/N_c$ 
included, and their coefficients resulting from 
numerical fits using the global wave function (\ref{WF}). 
The values of $c_i$ are indicated under the headings Fit n,
in each case.}
\label{operators}
\renewcommand{\arraystretch}{2} 
\begin{center}
{\tiny
\begin{tabular}{lrrr}
\hline
\hline
Operator \hspace{0cm} &\hspace{0.0cm} Fit 1 (MeV)  &\hspace{0cm} Fit 2 (MeV) &\hspace{0cm} Fit 3 (MeV) \\
\hline
$O_1 = N_c \ \1 $                            & $481 \pm5$   &  $484\pm4$  & $495\pm3$ \\
$O_2 = \ell^i s^i$                	     & $-31 \pm26$  & $3\pm15$ & $-30\pm 25$ \\
$O_3 = \frac{1}{N_c}S^iS^i$                  & $161\pm 16$  & $150\pm11$  & \\
$O_4 = \frac{1}{N_c}T^aT^a$                  & $169\pm36$  & $139\pm27$ & \\
$O_5 = \frac{15}{N_c}\ell^{(2)ij}G^{ia}G^{ja}$    & $-29\pm31$      &  & $-32 \pm 29$\\
$O_6 = \frac{3}{N_c}\ell^iT^aG^{ia}$            & $32\pm26$  &      & $28\pm 20 $\\
$O_7 =  \frac{3}{N_c^2} S^i T^a G^{ia}$ &  & & $649 \pm 61$ \\ 
\hline
$\chi_{\mathrm{dof}}^2$                                    & $0.43$            & $1.04$ & $0.24$ \\
\hline \hline
\end{tabular}}
\end{center}
\end{table}

In Table \ref{operators} we show results obtained with the exact wave 
function (\ref{WF}).
The first 6 operators are linearly independent. In SU(3), $O_7$
is also linearly independent, but not in SU(4). In the latter case  
it can be written as a linear combination of $O_1$, $O_3$ and $O_4$  
\cite{Galeta:2009pn}
\begin{equation}\label{O7}
O_7 = - \frac{3(4 N_c-9)}{16 N^3_c} O_1 
+ \frac{3(N_c-1)}{8 N_c} (O_3 + O_4).
\end{equation}
We did however include it in Table \ref{operators} but made two distinct
fits, either with $O_3$ and $O_4$ as independent operators (Fit 1 and Fit 2)
or with $O_7$ alone (Fit 3),
ignoring $O_3$ and $O_4$, as they are contained in $O_7$.
It turns out that $\chi_{\mathrm{dof}}^2$ is the best in 
the latter case. However the fit with independently varying
coefficient for $O_3$ and $O_4$ is  interesting for physical
reasons. Namely, in
Table \ref{MASSES1}, one can see that the 
isospin term $O_4$ becomes as dominant in $\Delta$ as the spin
term $O_3$ in $^4 N$ resonances, bringing in each case a contribution of about 
200 MeV to the mass. The fit of Table \ref{MASSES2}, with $O_7$ alone,
shows the same pattern, with a similar contribution. This gives more
physical insight into $O_7$ defined in Eq. (\ref{O7}).

\begin{table}[h!]
\caption{The partial contribution and the total mass $M$ (MeV), 
predicted by the $1/N_c$ expansion, Eq. (\ref{BMASS}),
using Fit 1 and the global wave function (\ref{WF}). 
The last two columns give  the empirically known resonance masses,
name and status.}\label{MASSES1}
\renewcommand{\arraystretch}{2}
\begin{center}
{\tiny
\begin{tabular}{lrrrrrrrrrr}\hline \hline
                    &      \multicolumn{6}{c}{Part. contrib. (MeV)}  & \hspace{0cm} Total (MeV)   & \hspace{0cm}  Exp. (MeV)\hspace{0cm}& &\hspace{0cm}  Name, status \hspace{.0cm} \\

\cline{2-7}
                    &   \hspace{-0.2cm}   $c_1O_1$  & \hspace{-0.3cm}  $c_2O_2$ & \hspace{-0.2cm}$c_3O_3$ &\hspace{-0.2cm}  $c_4O_4$ &\hspace{-0.3cm}  $c_5O_5$ &\hspace{-0.2cm} $c_6O_6$   &        \\
\hline
$^2N_{\frac{1}{2}}$        & 1444 & 10 & 40 & 42 & 0 & -8  &   $1529\pm 11$  & $1538\pm18$ & & $S_{11}(1535)$****  \\
$^4N_{\frac{1}{2}}$        & 1444 &  26 & 201& 42 & -31& -20 &   $1663\pm 20$  & $1660\pm20$ & & $S_{11}(1650)$**** \\
$^2N_{\frac{3}{2}}$        & 1444 & -5  & 40 & 42 & 0  &  4  &   $1525\pm 8$   & $1523\pm8$  & & $D_{13}(1520)$****\\
$^4N_{\frac{3}{2}}$        & 1444 & 10  & 201& 42 & 25 & -8 &   $1714\pm45$   & $1700\pm50$ & & $D_{13}(1700)$***\\
$^4N_{\frac{5}{2}}$        & 1444 & -16 & 201& 42  & -6 & 12 &   $1677\pm8$    & $1678\pm8$  & & $D_{15}(1675)$****\\
\hline
$^2\Delta_{\frac{1}{2}}$  &  1444 & -10  & 40 & 211 & 0  & -40   & $1645\pm30$  & $1645\pm30$ & & $S_{31}(1620)$**** \\
$^2\Delta_{\frac{3}{2}}$  &  1444 & 5  & 40 & 211 & 0  & 20   & $1720\pm50$  & $1720\pm50$ & & $D_{33}(1700)$**** \\ 
\hline \hline
\end{tabular}}
\end{center}
\end{table}

\begin{table}[h!]
\caption{
Same as Table \ref{MASSES1} but for Fit 3 of Table \ref{operators}.}
\label{MASSES2}
\begin{center}
{\tiny
\renewcommand{\arraystretch}{2}
\begin{tabular}{lrrrrrrrrr}\hline \hline
                    &      \multicolumn{5}{c}{Part. contrib. (MeV)}  & \hspace{0cm} Total (MeV)   & \hspace{0cm}  Exp. (MeV)\hspace{0cm}& &\hspace{0cm}  Name, status \hspace{.0cm} \\

\cline{2-6}
                    &   \hspace{-0.1cm}   $c_1O_1$  & \hspace{-0.1cm}  $c_2O_2$ & \hspace{-0.1cm}$c_5O_5$ &\hspace{-0.1cm}  $c_6O_6$ &\hspace{-0.1cm}  $c_7O_7$  &        \\
\hline
$^2N_{\frac{1}{2}}$        & 1486 & 10  & 0   & -7  & 41 &   $1529\pm 11$  & $1538\pm18$ & & $S_{11}(1535)$****  \\
$^4N_{\frac{1}{2}}$        & 1486 &  25 & -33 & -18 & 203&    $1663\pm 20$  & $1660\pm20$ & & $S_{11}(1650)$**** \\
$^2N_{\frac{3}{2}}$        & 1486 & -5  & 0   & 4   & 41  &     $1525\pm 7$   & $1523\pm8$  & & $D_{13}(1520)$****\\
$^4N_{\frac{3}{2}}$        & 1486 & 10  & 26  & -7  & 203 &    $1718\pm41$   & $1700\pm50$ & & $D_{13}(1700)$***\\
$^4N_{\frac{5}{2}}$        & 1486 & -15 & 7   & 11  & 203 &   $1677\pm8$    & $1678\pm8$  & & $D_{15}(1675)$****\\
\hline
$^2\Delta_{\frac{1}{2}}$  &  1486 & -10 & 0  & -35  & 203     & $1643\pm29$  & $1645\pm30$ & & $S_{31}(1620)$**** \\
$^2\Delta_{\frac{3}{2}}$  &  1486 & 5   & 0  & 18   & 203    & $1711\pm24$  & $1720\pm50$ & & $D_{33}(1700)$**** \\ 
\hline \hline
\end{tabular}}
\end{center}
\end{table}

\subsection{Discussion of the  decoupling scheme}

Here we examine the differences in the results obtained, on the one hand, with
the exact wave function (\ref{WF}) and, on the other hand, with the approximate 
wave function of Ref. \cite{CCGL}. The latter corresponds to the term
with $p = 2$ in Eq. (\ref{decex}). In the large $N_c$ limit
the amplitude  (\ref{2coreex}) can be neglected and  (\ref{2coreground}) 
can safely be taken equal to 1.

In the decoupling scheme one has
\begin{equation}
 S^i = s^i + S^i_c, \ \ T^a = t^a + T^a_c, \ \ G^{ia} = g^{ia} + G_c^{ia} 
\end{equation}
where the operators with a lower index $c$ act on the core  and
the lower case operators act on the decoupled quark.
Then in the SU(2)-spin Casimir operator $ S^i S^i = S^i_cS^i_c +
2s^i  S^i_c + s^i s^i$ the first two terms are considered linearly independent.
A similar treatment is made for $T^a T^a$. In addition,
because the matrix elements 
of $S^i_cS^i_c$ and $T^a_c T^a_c$ are identical for SU(4) symmetric spin-flavor states, 
$T^a_c T^a_c$ is ignored. The $s^i s^i$ and $t^a t^a$ are constants
which can be also be ignored. The linearly independent operators remain
$S^i_cS^i_c$, $s^i S^i_c$ and $t^a T^a_c$.  In Table \ref{defit1}
we show a fit which includes these operators plus $O_1$ and $O_2$. 
One can see that the fit is poor and that the coefficients $c_3$, 
$c_4$ and $c_5$ (the latter for the approximate wave function) 
obtain abnormally large values with alternating signs. This is a
clear indication that the linearly independence assumption is
not correct. 
 
\begin{table}[h!]
\caption{List of operators $O_i$ and their coefficients $c_i$ obtained 
in the numerical fit, Ref. \cite{MS1}, 
to the 7 known experimental masses of the lowest negative parity
resonances.   
} 
\label{defit1}
\renewcommand{\arraystretch}{1.25}
\begin{center}
{\tiny
 \begin{tabular}{lcc}
\hline \hline
$O_i$  &   \hspace{0cm}   $c_i$(MeV) with approx. w.f.   & \hspace{0cm} 
$c_i$(MeV) with  w.f. Eq. (\ref{decex}) \\ \hline
$O_1 = N_c \ \1 $    &   $  211 \pm 23$ &  $299 \pm 20$ \\
$O_2 = \ell^i s^i$  &  $3 \pm 15$  & $3 \pm 15$ \\
$O_3 = \frac{1}{N_c} s^iS_c^i $  & $-1486 \pm 141$ &  $-1096\pm 125$\\
$O_4 = \frac{1}{N_c} S_c^iS_c^i$ &  $1182\pm 74$ &  $1545 \pm 122$ \\
$O_5 = \frac{1}{N_c} t^aT_c^a$ &  $-1508\pm 149$  &  $417 \pm 79$ \\ 
\hline
$\chi_{\mathrm{dof}}^2$       & $1.56$ &  $1.56$      \\ \hline \hline 
\end{tabular}}
\end{center}
\end{table}

In Ref. \cite{MS1} we have considered the effect of  
$S^iS^i$ instead of its separate parts. The fit
is much better, the $\chi_{\mathrm{dof}}^2$ lowers to 1.04 and the values 
of the corresponding $c_i$ coefficients have natural sizes,
as defined, for example, in Ref. \cite{Pirjol:2003ye}.
A similar behaviour is observed  in the analysis of the isospin
operators.  Table \ref{defit4} shows
the result, which implies that the coefficients are
identical irrespective of the wave function, the exact (\ref{WF}),
or the approximate one from Ref. \cite{CCGL}. This is natural, 
because these operators are symmetric under 
any interchange of particles and it does not matter if the spin-flavor
state is fully symmetrized or not.

\begin{table}[h!]
\caption{Fit with the SU(2)-spin 
and SU(2)-isospin Casimir operators acting on the whole system. }\label{defit4}
\renewcommand{\arraystretch}{1.15}
\begin{center}
{\tiny
 \begin{tabular}{lcc}
\hline \hline
$O_i$  &   \hspace{0cm}  $c_i$(MeV)  & \hspace{0cm} 
$c_i$(MeV) \\
&  with  approx. w.f.  \hspace{0cm}  & with w.f. Eq. (\ref{decex}) \\ \hline
$O_1 = N_c \ \1 $    &   $484 \pm 4$ &  $484 \pm 4$ \\
$O_2 = \ell^i s^i$    &   $3 \pm 15$ &  $3 \pm 15$\\
$O'_3 = \frac{1}{N_c}\left(2 s^iS_c^i+S_c^iS_c^i+\frac{3}{4}\right)$ & $150 \pm 11$ &  $150 \pm 11$\\
$O'_5 = \frac{1}{N_c}\left(2 t^aT_c^a+T_c^aT_c^a+\frac{3}{4}\right) $ & $139 \pm 27$ & $139 \pm 27$  \\ \hline
$\chi_{\mathrm{dof}}^2$       & $1.04$ &  $1.04$ \\ \hline \hline
\end{tabular}}
\end{center}
\end{table}

The above study helped us to better understand the previous choice  
of operators in the scheme based on the 
separation of the system into a symmetric core of $N_c-1$ quarks
and an excited quark. 
There, as mentioned above, the SU(2)-isospin Casimir operator 
was written as $T^2 = T^2_c + 2 t \cdot T_c + 3/4$
and decomposed into 
three independent pieces, corresponding to the terms in the above decomposition. 
As mentioned, in SU(4) $T^2_c$ and $S^2_c$ have identical matrix elements 
because
the spin and isospin states of a symmetric core are identical, so that
$T^2_c$ was ignored. But $t \cdot T_c$ has different matrix
elements from $s \cdot S_c$ as one can clearly see from Table II
of Ref. \cite{CCGL}. Then in the decoupling scheme
the isospin can be introduced only through $t \cdot T_c$.
In Table \ref{defit1}
we have shown that the introduction of the operators 
$\frac{1}{N_c}t \cdot T_c$ together with  $\frac{1}{N_c}S^2_c$ 
and $\frac{1}{N_c}s \cdot S_c$ separately deteriorates the fit.
This may explain why $\frac{1}{N_c}t \cdot T_c$ has been avoided in previous 
numerical fits both in SU(4) \cite{CCGL} and in SU(6) 
\cite{Goity:2002pu}.

\section{Conclusions}
We plan to extend our study to SU(6), to look for differences 
between results obtained
with the exact wave function on the one hand,
and the approximate one, used in the decoupling scheme, on the other
hand.  The analytic procedure is described in Ref. \cite{MS4}. 
The SU(3)-flavor  Casimir operator is expected to bring new information 
with respect to SU(2)-isospin. 
This will complement 
another future work on the spectrum of both nonstrange and strange baryons
in the $1/N_c$ expansion, based on  the use of the matrix elements of
SU(6) generators derived in Ref. \cite{MS2}, needed for baryons in mixed 
symmetric states spin-flavor $[N_c-1,1]$.


\end{document}